# Fast Depth Imaging Denoising with the Temporal Correlation of Photons

Zhenchao Feng,[1] Weiji He,[1,*] Jian Fang,[1] Guohua Gu,[1] Qian Chen,[1] Ping Zhang,[3] Yuanjin Chen,[2] Beibei Zhou,[4] Minhua Zhou[4]

[1]Jiangsu Key Laboratory of Spectral Imaging & Intelligence Sense (SIIS), Nanjing University of Science and Technology, Nanjing, China, 210094

[2]East China Institute of Optoelectronic Integrated Devices, Bengbu, China, 233042

[3]Jiangsu North Huguang Optics & Electronics Co., Ltd., Wuxi 214000, China

[4]Nanjing University of Science and Technology, Nanjing 210094, China

[*]Corresponding author: hewj@mail.njust.edu.cn

*Abstract*—This paper proposes a novel method to filter out the false alarm of LiDAR system by using the temporal correlation of target reflected photons. Because of the inevitable noise, which is due to background light and dark counts of the detector, the depth imaging of LiDAR system exists a large estimation error. Our method combines the Poisson statistical model with the different distribution feature of signal and noise in the time axis. Due to selecting a proper threshold, our method can effectively filter out the false alarm of system and use the ToFs of detected signal photons to rebuild the depth image of the scene. The experimental results reveal that by our method it can fast distinguish the distance between two close objects, which is confused due to the high background noise, and acquire the accurate depth image of the scene. Our method need not increase the complexity of the system and is useful in power-limited depth imaging.

## I. Introduction

Time of flight (ToF) light detection and ranging (LiDAR) systems have been widely used for many applications including environmental monitoring, geological surveying, and underwater engineering [1-3]. Photon counting LiDAR uses Gm-APD (Geiger-mode Avalance Photo Diode) as the single-photon detector, which has the characteristic of single photon sensitivity and picosecond time response. The use of Gm-APD can greatly enhance the detection of the extremely weak signal, and acquire the depth image of large distance and high precision. For the depth imaging of LiDAR, it is typical to first build a photon-count histogram over time, then use a time-inhomogeneous Poisson process model to find a maximum likelihood estimate of scene depth [4], and finally apply a traditional image denoising algorithm. However, in the presence of high background noise, the echo signal is usually drowned in the noise, and the imaging accuracy of the maximum likelihood depth estimate degrades significantly.

Several methodologies of filtering out the false alarms generated by noise have been presented. Daniel G. Fouche and Markus Henriksson reported the probability model of the LiDAR using Gm-APD detectors [5-6]. They proposed an analysis of the detection probability and false-alarm probability for the detectors working in Geiger mode, which has provided a theoretical basis for further research. HongJin Kong developed a novel LiDAR system that was implemented by using two Gm-APDs with intensity dividing [7]. An AND gate is used to compare the electrical signals from the Gm-APDs, then the noise is filtered out. However, the energy of the laser-return pulse is divided in half, which results in the target detection probability decreasing significantly especially in the presence of strong background noise.

Zijing Zhang proposed a real-time noise filtering strategy that was called as the unit threshold method [8]. This method was implemented by dividing the Gm-APD array into many elementary units and using a threshold to filter out the noise. The use of Gm-APD array also results in the loss of the received laser-return pulse energy per pixel, which cannot be used in power-limited imaging. Apparently, this method is built in the aspect of system architecture and has increased the complexity of the system. Therefore, we start to research a method that can effectively filter out the noise by using imaging algorithm.

For taking a clear 3D image of the target in a short time, we propose a fast depth imaging denoising strategy based on the temporal correlation of signal photons. Detections generated by laser-return pulse have a strong temporal correlation in the time axis, which usually concentrate in the pulse width of emitted laser. On the contrary, detections generated by noise distribute dispersedly and randomly in the time domain. Based on this observation, we combine the mixture inhomogeneous Poisson probabilistic model with the temporal correlation of signal photons. By our method, it is capable of finding the correlative signal detections in the time axis and using the ToFs of correlative signal detections to reestablish the depth image of the scene.

## II. Imaging Model Analysis

The experimental 3D imaging LiDAR employing the denoising method proposed in this paper is shown in Fig. 1

[9]. A laser pulse with a wavelength of 830 nm is emitted by the pulsed laser source and passes through the X/Y scanning mirrors. The laser-return pulse and background light are collected by the optical system, and then trigger the detector of Gm-APD that has a dead time of 50ns and dark count of fewer than 100 counts per second. The response of Gm-APD is recorded by the TCSPC (time-correlated single-photon counting) module with 4 ps minimum time-bin width. The computer is used to coordinate the operation of different system parts. The ToF of each photon detection event and the number of emitted laser pulse are recorded at every image pixel.

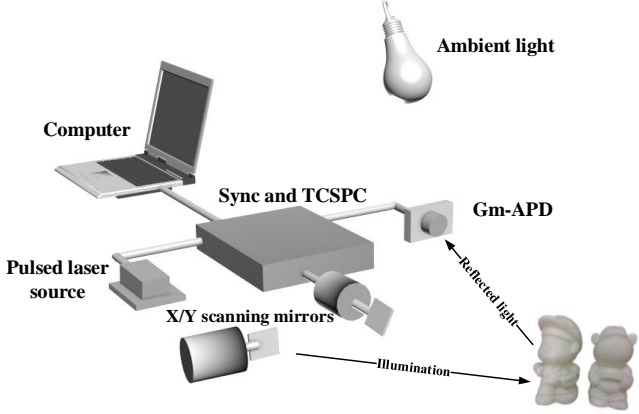

Fig. 1. 3D imaging LiDAR.

**2.1 Probability Analysis**

Define $b$ as the number of time bins within the range gate $T_{gate}$. Assume that the total number of photon counts generated by background noise and dark current $N$ is constant during the data acquisition. Then the noise photon counts distributing in each time bin are $n = N/b$. Define $S$ as the total photon counts generated by laser-return pulse. Define $g$ as the serial number of the target time bin, which can be approximatively measured before the data acquisition. Thus, the probabilities of signal and noise detections are [5]:

$$P_{sig} = \exp(-gN) \times \{1 - \exp(-S-n)\}. \quad (1)$$
$$P_{noi} = 1 - P_{sig} - \exp(-S-N). \quad (2)$$

Ignoring the effect of the laser-return pulse broadening, the signal detections possibly appear only during the repetition period of laser pulse $T_f$ ($T_f = 400$ ns in our experiments), and mainly concentrate on the pulse width $T_p$ ($T_p = 200$ ps in our experiments). The signal photon counts are characterized by the short-duration illumination pulse [10], so the ToFs for signal counts have a small variance. Thus, the criterion of finding out the correlative signal detections is:

$$|T_1 - T_2| \leq T_p \quad (3)$$

Wherein, $T_1$ and $T_2$ are the ToFs (time-of-fight) of two detections, respectively. For the detection probability of the correlative signal and noise, there are four interesting cases to consider, as follows:

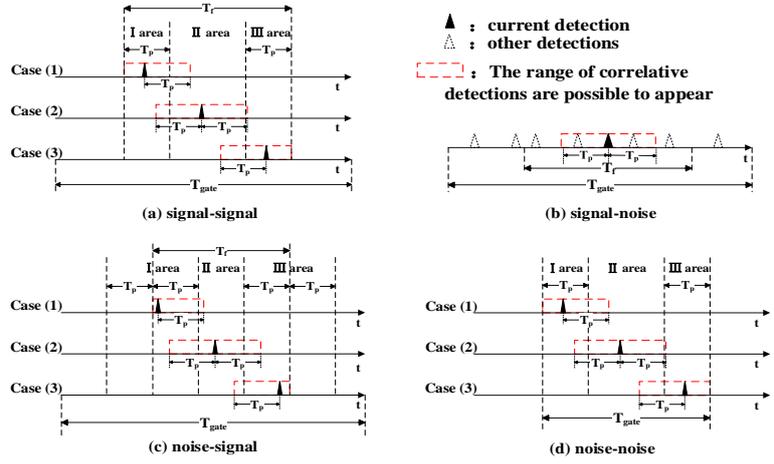

Fig. 2. Detection probability of the correlative signal and noise.

*A. The current detection is signal, and the next detection is signal.*

The time duration $T_f$ is divided into $T_f/\tau$ time bins, where $\tau$ is the width of each time bin. As shown in Fig.2 (a), only when the current detection is distributed in any areas of Ⅰ, Ⅱ, and Ⅲ, it is possible to have temporal correlation with the next detection. The probability of the signal detection distributed in each time bin is $\tau/T_f$. Thus, the probability $P_{ss}$ that the current signal detection has a temporal correlation with the next signal detection is:

$$P_{ss} = P_{sig} \times \frac{\tau}{T_f} \times [\underbrace{(\frac{T_p}{T_f} + \frac{T_p + \tau}{T_f} + \ldots + \frac{2T_p}{T_f})}_{\text{I area}} + \underbrace{(\frac{T_f}{\tau} - \frac{2T_p}{\tau} - 2) \times \frac{2T_p}{T_f}}_{\text{II area}}$$
$$+ \underbrace{(\frac{2T_p}{T_f} + \frac{2T_p - \tau}{T_f} + \ldots + \frac{T_p}{T_f})}_{\text{III area}}] \quad (4)$$
$$= P_{sig} \times \frac{\tau}{T_f} \times [(\frac{T_p}{\tau} + 1) \times \frac{1.5T_p}{T_f} \times 2 + (\frac{T_f}{\tau} - \frac{2T_p}{\tau} - 2) \times \frac{2T_p}{T_f}]$$
$$\approx P_{sig} \times \frac{2T_p T_f - T_p^2}{T_f^2}$$

*B. The current detection is signal, and the next detection is noise.*

As shown in Fig.2 (b), when the time duration between the current signal detection and the next detection is in the range of $T_p$, the above-mentioned two detections are correlative. The probability that the next detection is noise is $P_{noi}$. Thus, the probability that the current signal detection has a temporal correlation with the next noise detection is:

$$P_{sn} = P_{noi} \times \frac{2T_p}{T_{gate}} \quad (5)$$

*C. The current detection is noise, and the next detection is signal.*

As shown in Fig.2 (c), only when the current detection is distributed in any areas of Ⅰ, Ⅱ, and Ⅲ, it is possible to have a temporal correlation with the next detection. These three areas are divided into $(T_f + 2T_p)/\tau$ time bins. The probability that the current noise detection is distributed in each time bin is $\tau/T_{gate}$, the probability that the next detection is signal is $P_{sig}$. Thus, the probability $P_{ns}$ that the current noise detection has a temporal correlation with the next signal detection is:

$$P_{ns} = P_{sig} \times \frac{\tau}{T_{gate}} \times [\underbrace{(\frac{\tau}{T_f} + \frac{2\tau}{T_f} + ... + \frac{2T_p - \tau}{T_f})}_{\text{I area}} + \underbrace{(\frac{T_f + 2T_p}{\tau} - \frac{4T_p}{\tau} + 2) \times \frac{2T_p}{T_f}}_{\text{II area}}$$

$$+ \underbrace{(\frac{2T_p - \tau}{T_f} + \frac{2T_p - 2\tau}{T_f} + ... + \frac{\tau}{T_f})}_{\text{III area}}]$$

$$= P_{sig} \times \frac{\tau}{T_{gate}} \times [(\frac{2T_p}{\tau} - 1) \times \frac{T_p}{T_f} \times 2 + (\frac{T_f + 2T_p}{\tau} - \frac{4T_p}{\tau} + 2) \times \frac{2T_p}{T_f}]$$

$$\approx P_{sig} \times \frac{2T_p}{T_{gate}}$$

(6)

*D. The current detection is noise, and the next detection is noise.*

This case is similar to case A as shown in Fig.2 (d). Only when it is distributed in the areas of Ⅰ, Ⅱ, and Ⅲ, the current detection is possible to have a temporal correlation with the next detection. The probability of the noise detection distributed in each time bin is $\tau / T_{gate}$. The probability $P_{nn}$ that the current noise detection has a temporal correlation with the next noise detection is:

$$P_{nn} = P_{noi} \times \frac{2T_p T_{gate} - T_p^2}{T_{gate}^2} \quad (7)$$

The use of a judgmental window in the timeline and a threshold has proved to be a simple and useful way for separating signal and noise [11-12]. Our purpose is to find out the correlative signal detections in the time axis per pixel. We use a judgmental range moving along the time axis to find the correlative signal detections as shown in Fig. 2 and Fig. 3. Each photon detection event is an independent process, so the probability $P_{ssc}$ that at least $K$ signal detections are correlative is:

$$P_{ssc} = \sum_{m=K}^{M} P_{ss}^m = \sum_{m=K}^{M} (P_{sig} \times \frac{2T_p T_f - T_p^2}{T_f^2})^m \quad (8)$$

, where $M$ is the number of detections within the judgmental range. The false-detection probability $P_{nnc}$ of our method is:

$$P_{nnc} = \sum_{m=K}^{M} (P_{sn} + P_{ns} + P_{nn})^m$$

$$= \sum_{m=K}^{M} (P_{noi} \times \frac{2T_p}{T_{gate}} + P_{sig} \times \frac{2T_p}{T_{gate}} + P_{noi} \times \frac{2T_p T_{gate} - T_p^2}{T_{gate}^2})^m.$$

(9)

**2.2 Strategy**

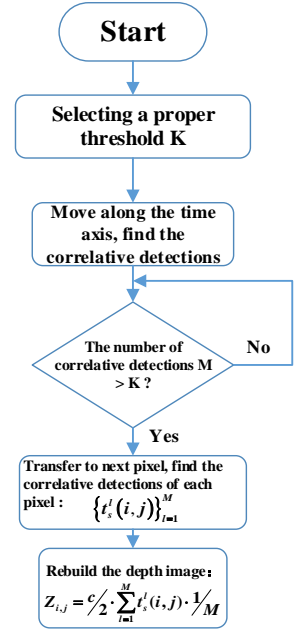

Fig. 3. Flow diagram of our strategy.

The flow diagram of our strategy is shown in Fig. 3. Selecting a proper threshold K is the first step of our strategy. According to Fig. 4, the threshold K has a direct effect on the value of $P_{ssc}$ and $P_{nnc}$. In order to eliminate the influence of noise, the selection of a proper threshold should try to make $P_{ssc}$ is the largest and $P_{nnc}$ is the minimum. However, the reflected signal intensity of each pixel is different and unknown. According to Fig. 4(a), the probability of correlative signal detections $P_{ssc}$ is decreased with the increasing of the threshold K at the same signal intensity. Thus, we should choose a small K. We set a warning line $f$ of $P_{nnc}$ to limit the interference of the noise detections. Therefore, the criteria of selecting a proper threshold are: 1) the threshold K should ensure $P_{nnc}$ below the warning line $f$; 2) under the condition of satisfying the above 1), it is better to choose a smaller K. According to the intensity of background noise, which is measured before the data acquisition, the numerical results of selecting a proper threshold are shown in Fig. 4(c).

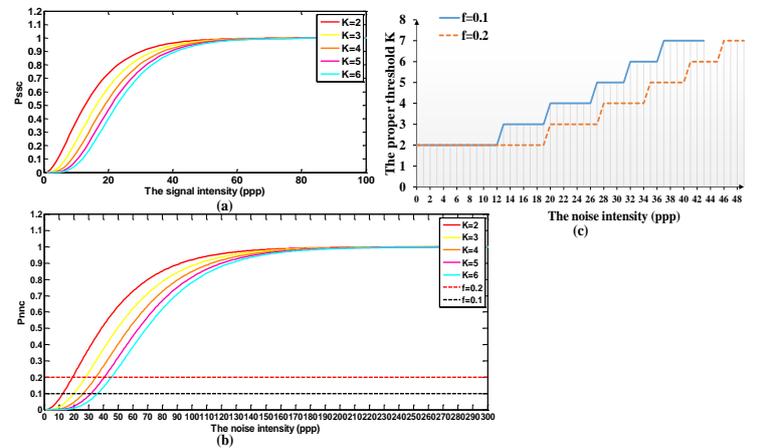

Fig. 4. As shown in herein (a) and (b), there are different probabilities of $P_{ssc}$ and $P_{nnc}$ (the vertical axis) with different signal and noise intensities (the horizontal axis, photons per pixel (ppp) ). In order to effectivity filter out the noise, a proper threshold K is selected in advance as shown in herein (c).

Define the obtained ToF dataset in the pixel $(i,j)$ as $\{t^l(i,j)\}_{l=1}^n$, where $n$ is the number of detections in $(i,j)$. We use the detection $t^l(i,j)$ as the center, and find out the number of detections within the scope of $T_p$ before and after this detection as shown in Fig. 2. According to Eq. (3), the ToFs of the correlative detections of $t^l(i,j)$ are:

$$\{t^M(i,j)\in[t^l(i,j)-T_p,\ t^l(i,j)+T_p],\ 1\le l\le n\}. \quad (10)$$

As shown in Fig.3, if the number of these correlative detections $M$ is smaller than the threshold $K$, we continue to use the next detection as the center and find out the correlative detections of the next detection; If the number of these correlative detections $M$ is larger than the threshold $K$, these detections are classified as correlative signal detections. The ToFs of these correlative signal detections are used as the depth estimation in the pixel $(i,j)$. Then, we transfer to the next pixel until we find out the correlative signal detections of each pixel $\{t_s^l(i,j)\}_{l=1}^M$. Ultimately, we obtain the depth image of the scene using the average ToFs of the correlative signal detections:

$$Z_{i,j} = c/2 \cdot \sum_{l=1}^M t_s^l(i,j) \cdot 1/M \quad (11)$$

### III. EXPERIMENT RESULTS AND ANALYSIS

The experimental scene is shown in Fig. 5(a). The distance between two objects is 10cm, and the distance between the latter object and the wall is also 10cm. The distance between the experimental scene and the LiDAR system is 20m. A daylight lamp is used to simulate the solar background environment, and the signal-background-ratio is that $SBR=1$. The size of obtained depth image is $300\times300$ pixels. We use RMSE (root mean-square error) as the evaluation criterion of depth estimation:

$$RMSE(z,z') = \sqrt{\frac{1}{n^2}\sum\sum(z-z')^2} \quad (12)$$

, where $z$ is the real depth value and $z'$ is the depth estimation value.

The obtained raw data is shown in Fig.5 (b). For decreasing the interference of noise as much as possible, the warning line of $P_{nc\_window}$ is set as $f=0.2$. The noise intensity is 25 photons per pixel. According to Fig 4(c), we set the threshold $K=3$ in our experiment. The experimental results of applying our method are shown in Fig. 5(c) and 5(d). Fig. 5(c) and 5(d) are the same processed results shown in the different angles of view.

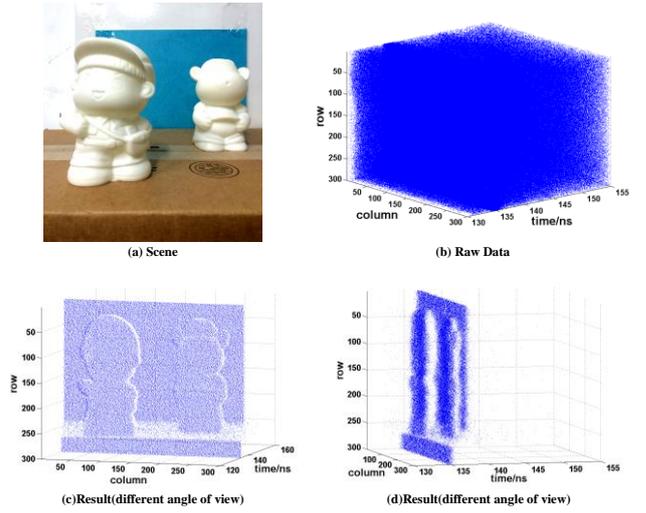

Fig. 5. Experimental results.

As shown in Fig. 5(b), due to the high noise environment, any point in time axis is possible to have photon detections arising. The detections generated by the laser-return pulse are drowned in the noise detections. According to Fig. 5(c) and 5(d), it's intuitively shown that the detections generated by background noise can be effectively filtered out by our method and the detections generated by the laser-return pulse are found out.

The denoising results obtained by our method and the depth estimation based on maximum likelihood (MLE) are compared in Fig 6. Fig. 6(a) is the ground truth depth image of the experimental scene. Fig. 6(b) is the depth image obtained by the maximum likelihood depth estimation and median filter. Fig. 6(c) is the image of the absolute error between the maximum likelihood depth estimation and ground truth. Fig. 6(d) and 6(e) are the depth image and the absolute error image of applying our method, respectively.

Table 1. Comparison of imaging accuracy and dwell time

|  | SBR=1 | | SBR=10 | |
| --- | --- | --- | --- | --- |
|  | RMES/m | Dwell time/ms | RMES/m | Dwell time/ms |
| MLE | 0.3851 | 0.7523 | 0.2062 | 0.7136 |
| Our Method | 0.0487 | 0.1094 | 0.0364 | 0.0533 |

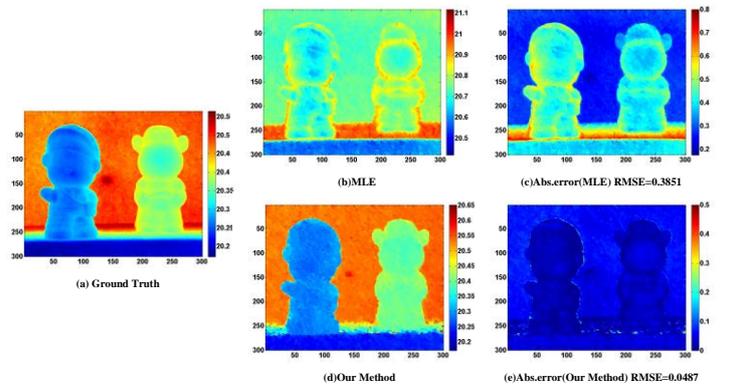

Fig. 6. Comparison between our method and the maximum likelihood depth estimation in the condition of SBR=1.

As shown in Fig. 6(b) and 6(c), there is a large depth imaging error of applying the maximum likelihood depth estimation, and the distance between two objects is hardly discriminated. The reason is that the background noise is too strong, resulting in the noise detections arising in a long time range. Even though it has spent a long dwell time in every pixel, there is a large error of applying the maximum likelihood depth estimation. As shown in Fig. 6(d) and 6(e), by our method, it is capable of discriminating the distance of two objects, and the ultimate depth image is close to the ground truth. Comparing to the maximum likelihood depth estimation, the imaging accuracy of applying our method has been increased by 8-fold. As shown in Table 1, the dwell time of the depth estimation based on maximum likelihood is 7 times longer than our method. The acquisition time of our method is shorter, because it only uses the ToFs of the correlative signal detections to rebuild the depth image. Meanwhile, when the number of correlative detections is satisfied with the threshold K, it will immediately transfer to the next pixel. We have repeated the above experiment but in the condition of SBR=10. As shown in Table.1, comparing to MLE, the imaging accuracy of applying our method has been increased by 5-fold in the condition of SBR=10. Since there is more signal in the condition of SBR=10, our method can more easily find out the signal detections and transfer to the next pixel. Thus, the dwell time of applying our method in the condition of SBR=10 is shorter than that in the condition of SBR=1 as shown in Table.1. In brief, our method has a good real-time character and imaging accuracy in the high background light environment.

For demonstrating the effect of threshold K, we have carried out another experiment. The experimental scene that an object is placed in front of the wall at a distance of 5cm is depicted in Fig. 7(a). The background environment of this experiment is $SBR=1$. The non-processed raw data are shown in Fig. 7(b). And the results obtained by different thresholds are shown in Fig. 7(c)~(g).

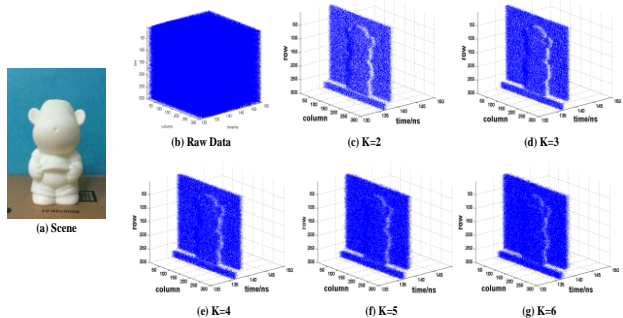

Fig. 7. Results of employing different threshold.

Table 2. Effect of the threshold K

|  | K=2 | K=3 | K=4 | K=5 | K=6 | MLE |
|---|---|---|---|---|---|---|
| **RMSE**/m | 0.0245 | 0.0232 | 0.0300 | 0.0366 | 0.0367 | 0.2369 |
| **Dwell time** /ms | 0.0556 | 0.1501 | 0.1040 | 0.1269 | 0.1601 | 1.5124 |

As shown in Fig. 7 and Table 2, the imaging accuracy is slightly decreased with the increasing of the threshold K. The main reason is that the signal detections in the judgmental range are mixed with the noise detections due to selecting an improper threshold. Because the number of the detections that are used for depth image reconstruction is increased with the increasing of the threshold, the dwell time also becomes longer. When the threshold $K$ is equal to 3, the depth image obtained by our method is the closest to the ground truth, which is in accordance with the results of the theoretical derivation.

## IV. CONCLUSIONS

In this work, we propose a fast depth imaging denoising strategy based on the temporal correlation of laser-return photons. Our method combines the different distribution feature of signal detections and noise detections in the time axis with the Poisson statistical model, and it is capable of distinguishing the signal photons between noise photons. As the noise detections are filtered out by our method, it is able to obtain a more accurate depth estimation of the scene. Since we only use the ToFs of signal detections, the depth image of employing our method is obtained in a short dwell time. Comparing to traditional depth image processing method, by our method, it is capable of distinguishing the distance between the close objects in the existence of strong background noise. The image accuracy of applying our method is increased by 8-fold, and the dwell time of obtaining the depth image is also 7 times shorter than the traditional method. Our method expands the application of LiDAR in the high background light environment and is also useful in power-limited imaging.

ACKNOWLEDGMENT

The authors gratefully acknowledge the supports from the Seventh Six-talent Peak project of Jiangsu Province (Grant No. 2014-DZXX-007), the National Natural Science Foundation of China (Grant No. 61271332), the Fundamental Research Funds for the Central Universities (Grant No. 309920140112012), the Innovation Fund Project for Key Laboratory of Intelligent Perception and Systems for High-Dimensional Information of Ministry of Education (Grant No. JYB201509), and the Fund Project for Low-light-level Night Vision Laboratory (Grant No. J20130501).